\documentclass[a4paper,12pt]{article}

\usepackage{graphics}

\begin{document}

\title{
\begin{flushright}
{\normalsize IRB-TH-7/03}
\end{flushright}
\vspace{2 cm}
\bf Partial rip scenario - \\
a cosmology with a growing  \\
cosmological term 
}

\author{ H. \v Stefan\v ci\'c\thanks{shrvoje@thphys.irb.hr}
}

\vspace{3 cm}
\date{
\centering
Theoretical Physics Division, Rudjer Bo\v{s}kovi\'{c} Institute, \\
   P.O.Box 180, HR-10002 Zagreb, Croatia}


\maketitle

\abstract{A cosmology with the growing cosmological term is
considered. If there is no exchange of energy between vacuum and matter
components,
the requirement of general covariance implies the time
dependence of the gravitational constant $G$. Irrespectively of the exact
functional form of the cosmological term growth, the universe ends in a
de Sitter regime with a constant asymptotic $\Lambda$, but vanishing $G$. 
Although there is no divergence of
the scale factor in finite time, such as in the ``Big Rip" scenario, 
gravitationally
bound systems eventually become unbound. In the case of systems bound by
non-gravitational forces, there is no unbounding effect, as the asymptotic 
$\Lambda$ is insufficiently large to disturb these systems.}

\vspace{2cm}

Cosmological observations of increasing quantity, quality and diversity have
established a new picture of the universe, its composition and dynamics
\cite{Rev}.
Measurements of luminosity-redshift relations for the supernovae of type Ia
(SNIa) \cite{SN} and the temperature anisotropies of the cosmic microwave
background radiation (CMBR) \cite{CMBR}, in agreement with other 
cosmological observations,
have determined that the universe is presently in the phase of accelerated
expansion, which is attributed to a new component of the universe named {\em dark
energy}. The origin of dark energy (which together with {\em dark matter} 
constitutes the ``dark sector" of the universe) has not yet been clarified.
Pretenders to the title of dark energy are, however, numerous. The first
in a long line of candidates is certainly the cosmological constant (CC),
also known as cosmological term \cite{wein,peeb,pad}. 
This conceptually simple choice, which naturally fits into the
formalism of general relativity (GR), is, however, burdened with problems 
which, to be solved, require a highly unnatural amount of fine-tuning. Namely,
quantum field theory contributions to the vacuum energy (which is equivalent to
the CC) and the observed value of the CC differ by many orders of magnitude.
Furthermore, there is a coincidence problem of understanding why the CC
energy density and the non-relativistic matter energy density are 
comparable at the
present epoch. These problems were a strong incentive towards the development of
dynamical dark energy models. One of the most studied classes of such models
are {\em quintessence} models \cite{Q}. Quintessential models describe
dark energy in terms of the scalar field slowly rolling in a potential. Another
appealing proposal is the Chaplygin gas \cite{CG}. The most interesting feature
of this model is the possibility of unified description of the 
``dark sector", i.e. the unification of dark matter and dark energy.
Another alternative to quintessence with the similar properties of
effective unification of dark sector is Cardassian cosmology \cite{Freese}.
Other interesting models include the non-perturbative effects of vacuum energy
 \cite{Parker} and the fluctuating CC \cite{Ahmed}. 
The consideration of the CC as a dynamical quantity in the framework of the
renormalization group equation \cite{mi,sola,bonanno} is another promising
approach.

A large majority of dark energy models describes dark energy in terms of the
equation of state (EOS)
\begin{equation}
\label{eq:eos}
p_{d} = w \rho_{d} \, ,
\end{equation}
where $w$ is the parameter of the EOS, while $p_{d}$ and $\rho_{d}$ denote the
pressure and the energy density of dark energy, respectively. The value
$w=-1$ is characteristic of the cosmological constant, while the dynamical
models of dark energy generally have $w \ge -1$. A recent examination of the
dark energy EOS, based on the data from CMBR, SNIa, large scale structure (LSS)
and Hubble parameter measurements from the Hubble Space Telescope (HST),
assuming the redshift independent parameter $w$, restrict $w$ to be in the
interval $-1.38 < w < -0.82$ at the 95\% confidence level \cite{odman}. This
result raises a question about the possibility of dark energy models with the
supernegative EOS, i.e. with $w < -1$. This new sort of dark energy, first
analyzed in \cite{cald} which was followed by numerous analyses \cite{phantom}, 
has soon deserved a term of its own, 
{\em phantom energy}. The most interesting feature of phantom energy models
is the possibility of a divergence of the scale factor of the universe $a$
in finite
time. Such a behaviour of the scale factor has a dramatic effect on all
bound systems. Namely, the bound systems become unbound at some finite time
interval before the onset of the divergence in $a$. This type of the fate of the
universe is known as a ``Big Rip" scenario \cite{cald2}.

In this letter we consider the case of the growing cosmological term 
$\Lambda$ and its implications for the asymptotic expansion of the universe and
the destiny of the bound systems. Let us start with the specification of the main
characteristics of the components of the universe. We assume that we have
two components of the universe: non-relativistic matter and the
variable cosmological term. The inclusion of radiation 
or other components is straightforward, but
without any substantial influence on the future evolution of the universe.
Non-relativistic matter has the EOS
\begin{equation}
\label{eq:nonrel}
p_{m} = \gamma \rho_{m} \, ,
\end{equation}
where $\gamma \ge 0$ is the parameter of the EOS and $p_{m}$ and $\rho_{m}$ 
represent
the pressure and the energy density of non-relativistic matter, respectively.
The conservation of the energy-momentum tensor of non-relativistic matter,
$T^{\mu \nu}_{m ; \nu} = 0$, leads to the standard relation of the evolution of
$\rho_{m}$ with the scale factor
\begin{equation}
\label{eq:rhononrel}
\rho_{m} = \rho_{m,0} \left( \frac{a}{a_{0}} \right)^{-3 (1+\gamma)} \, .
\end{equation} 
The equation of state of the cosmological term is 
\begin{equation}
\label{eq:lambda}
p_{\Lambda} = - \rho_{\Lambda} \, ,
\end{equation}
where $p_{\Lambda}$ is the pressure and $\rho_{\Lambda}$ is the energy density of
the cosmological term. 
The requirement of the conservation of the energy-momentum tensor of
cosmological term,
$T^{\mu \nu}_{\Lambda} = \rho_{\Lambda} g^{\mu\nu}$, leads to a constant 
cosmological term
energy density $\rho_{\Lambda}$. If we, however, consider a variable
(time-dependent) $\rho_{\Lambda}$, $T^{\mu \nu}_{\Lambda}$ can no longer be
conserved. Models with variable $\Lambda \equiv 8 \pi G \rho_{\Lambda}$ 
and $G$ were extensively studied in \cite{Gvar}. The consideration of phantom
energy models in an analogous framework leads to an interesting general finding on
the future evolution of the universe \cite{stef}.
In the setting described above, the general covariance of the Einstein equation
\begin{equation}
\label{eq:Einstein}
G_{\mu \nu} = -8 \pi G T_{\mu \nu} \, , 
\end{equation}   
where $G_{\mu \nu}$ is the Einstein tensor and $T^{\mu \nu} = T^{\mu \nu}_{m} +
T^{\mu \nu}_{\Lambda}$, can be maintained if the gravitational constant $G$
acquires space-time dependence. This fact can be interpreted as a modification
of the dynamics of General Relativity. We effectively describe this 
additional dynamics by promoting $G$ into a function of space-time coordinates.   
We further assume that $G$ depends on time only, i.e. $G
= G(t)$ \footnote{Interesting models with variable $G$ are also obtained in the
formalism of Quantum field theory in the curved space-time \cite{Odin}.} . 
The condition $(G(t)T^{\mu \nu})_{;\nu} = 0$ can be expressed as
\begin{equation}
\label{eq:Gevol1}
d(G(\rho_{m}+\rho_{\Lambda}) a^{3}) = -G (p_{m}+p_{\Lambda})d a^{3} \, , 
\end{equation} 
which gives the law of evolution of $G(t)$
\begin{equation}
\label{eq:Gevol2}
\dot{G}(\rho_{m}+\rho_{\Lambda}) + G \dot{\rho_{\Lambda}} = 0 \, . 
\end{equation}

Apart from the procedure displayed above, there exist other ways how the
variability of the cosmological term can be incorporated into the formalism of
General Relativity. One possibility is to allow the interchange of energy and
momentum between the cosmological term (vacuum) and matter and radiation
components \cite{FAFM}. In this model, the energy-momentum tensors of
separate components (non-relativistic matter or radiation and vacuum energy)
are not conserved, but the total energy-momentum tensor is. Furthermore, 
$G$ is time-independent in this setting. It is interesting to see
that both the model of our paper and the model given in \cite{FAFM} follow from
equation (\ref{eq:Gevol1}) under different assumptions. The dynamics of our
model (\ref{eq:Gevol2}) is obtained assuming that the vaccum energy is time
dependent and that the tensor of energy-momentum of the matter component is
separately conserved. The dynamics of \cite{FAFM}, expressed by their equation 
(5a), is recovered under the assumption that $G$ is time-independent and that
there exists an exchange of energy between the vacuum and the matter components. 

The set of equations governing the evolution of the universe is completed by
the Friedmann equations for the scale factor
\begin{equation}
\label{eq:Fried1}
\left( \frac{\dot{a}}{a} \right)^{2} + \frac{k}{a^{2}} = \frac{8 \pi}{3} G
(\rho_{m}+\rho_{\Lambda})  \, , 
\end{equation}
\begin{equation}
\label{eq:Fried2}
\frac{\ddot{a}}{a} = -\frac{4 \pi}{3} G (\rho_{m} + \rho_{\Lambda} + 3p_{m} +
3p_{\Lambda}) \, . 
\end{equation}

Given the set of equations (\ref{eq:rhononrel}), (\ref{eq:Gevol2}), 
(\ref{eq:Fried1}) and (\ref{eq:Fried2}), we can investigate the future evolution
of the universe for a general growing $\rho_{\Lambda}$. The law of evolution of
$\rho_{m}$ (\ref{eq:rhononrel}) clearly shows that at sufficiently distant
future times we have $\rho_{\Lambda} \gg \rho_{m}$. In this limit, equations
(\ref{eq:Gevol2}) and (\ref{eq:Fried1}) become
\begin{equation}
\label{eq:distant1}
\dot{G} \rho_{\Lambda} + G \dot{\rho_{\Lambda}} = 0 \, , 
\end{equation}
\begin{equation}
\label{eq:distant2}
\left( \frac{\dot{a}}{a} \right)^{2} + \frac{k}{a^{2}} = \frac{8 \pi}{3} G
\rho_{\Lambda}  \, . 
\end{equation}
From (\ref{eq:distant1}) it follows that at distant times we have
$G \rho_{\Lambda} = const$. Therefore, in equation (\ref{eq:distant2}) we
can disregard the $k/a^{2}$ term at sufficiently distant times. 
We finally obtain the equations governing the evolution of the universe
\begin{equation}
\label{eq:govevol1}
\frac{d (G \rho_{\Lambda})}{dt} = 0 \, , 
\end{equation}
\begin{equation}
\label{eq:govevol2}
\left( \frac{\dot{a}}{a} \right)^{2}  = \frac{8 \pi}{3} G \rho_{\Lambda}  \, . 
\end{equation}
The solution of the equation displayed above is of the form
\begin{equation}
\label{eq:solasym}
a \sim e^{\sqrt{\frac{\Lambda_{\infty}}{3}} t} \, ,
\end{equation}
where $\Lambda_{\infty} = 8 \pi (G \rho_{\Lambda})_{t \rightarrow \infty}$.
In the distant future the universe enters the de Sitter regime of expansion.
This fact does not depend on the concrete form of 
$\rho_{\Lambda}$, as long as it is a growing function of time (explicitly or
implicitly via some other quantity depending on time, such as the scale factor
$a$). Furthermore, the growth
of $\rho_{\Lambda}$ implies that $G$ is a decreasing function at distant times.
The unbounded growth of $\rho_{\Lambda}$ results in vanishingly small values of 
$G$ at distant times. 

Let us further examine the complete evolution of the universe for a class 
of growing $\rho_{\Lambda}$ models. We assume the following form for the
variable cosmological term energy density:
\begin{equation}
\label{eq:varLambda}
\rho_{\Lambda} = \rho_{\Lambda,0} \left( \frac{a}{a_{0}}\right)^{-3 (1+\eta)} \, ,
\end{equation}
where $\eta < -1$. We consider the flat universe case, $k = 0$, which implies
that the total energy density $\rho_{0}$ equals the critical energy density 
$\rho_{c,0}$, and introduce a
dimensionless parameter $\Omega_{\Lambda}^{0} \equiv \rho_{\Lambda,0}/\rho_{0}$.
The subscript or superscript $0$ denotes the present time throughout the paper.
 What remains is
solving equations (\ref{eq:Gevol2}) and (\ref{eq:Fried1}) given the laws of
evolution of the energy densities (\ref{eq:rhononrel}) and (\ref{eq:varLambda}).  
The quantity $G$ evolves according to the law
\begin{equation}
\label{eq:Glaw}
G = G_{0} \left( \Omega_{\Lambda}^{0} \left( \frac{a}{a_{0}}
\right)^{-3(\eta-\gamma)} + 1 - \Omega_{\Lambda}^{0}
\right)^{-\frac{1+\eta}{\eta-\gamma}}   \, ,
\end{equation} 
while the scale factor is implicitly given by the expression
\begin{equation}
\label{eq:alaw}
H_{0} (t-t_{0}) = \int_{1}^{a/a_{0}} x^{\frac{1}{2}(1+ 3 \gamma)} \left( 
\Omega_{\Lambda}^{0} x^{-3(\eta-\gamma)} + 1 - \Omega_{\Lambda}^{0}
\right)^{\frac{1}{2} \frac{1+\gamma}{\eta-\gamma}} dx \, .
\end{equation}

\begin{figure}
\centerline{\resizebox{1.0\textwidth}{!}{\includegraphics{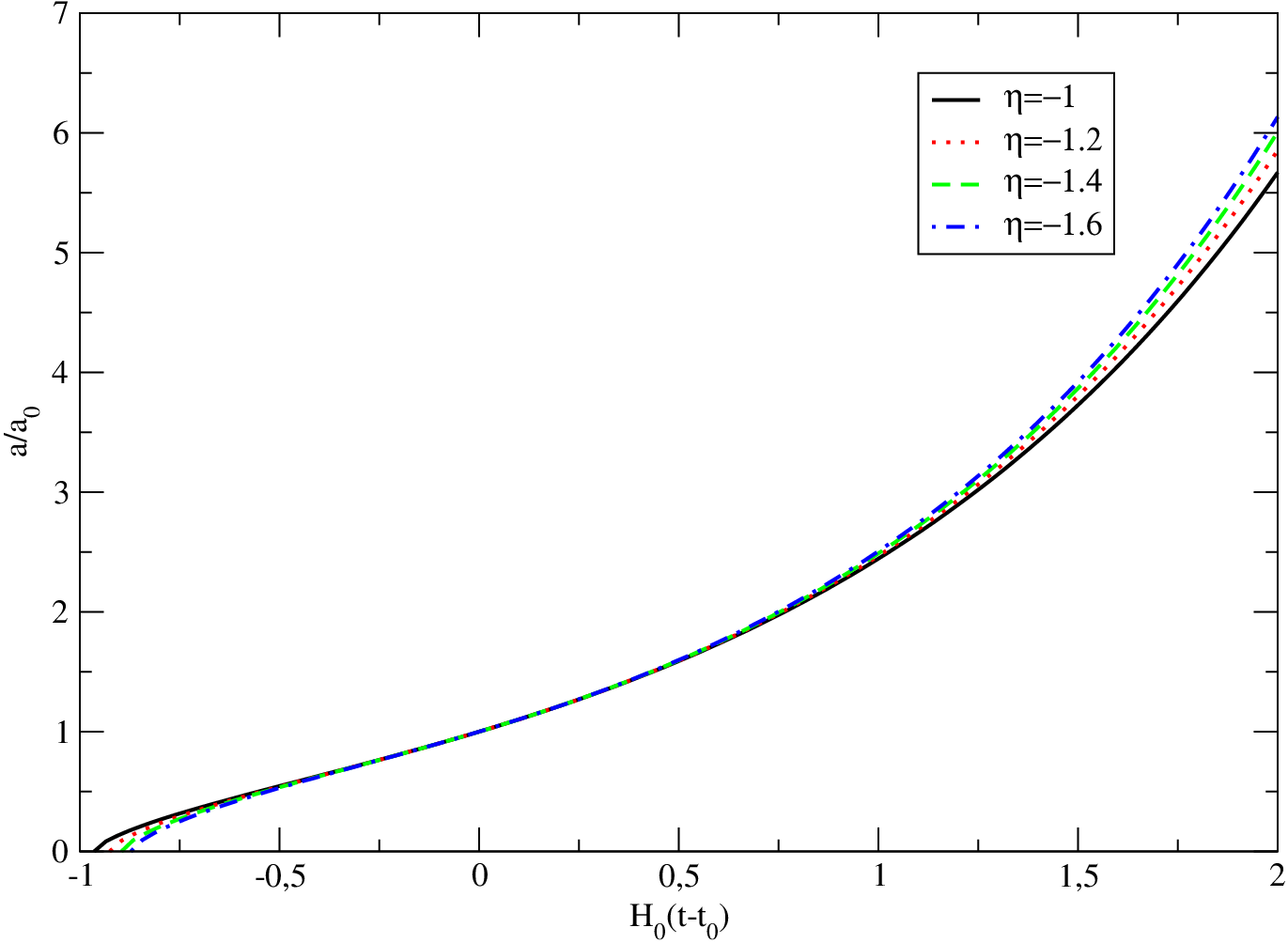}}}
\caption{\label{fig:a} The evolution of the scale factor of the universe is
shown for $\Omega_{\Lambda}^{0} = 0.7$, $\gamma = 0$ and four different values of 
the parameter $\eta$. The difference between
various models is the most prominent in the distant past and the distant future.
More negative values of $\eta$ lead to the faster growth of the scale factor in
the distant future.
}
\end{figure}

\begin{figure}
\centerline{\resizebox{1.0\textwidth}{!}{\includegraphics{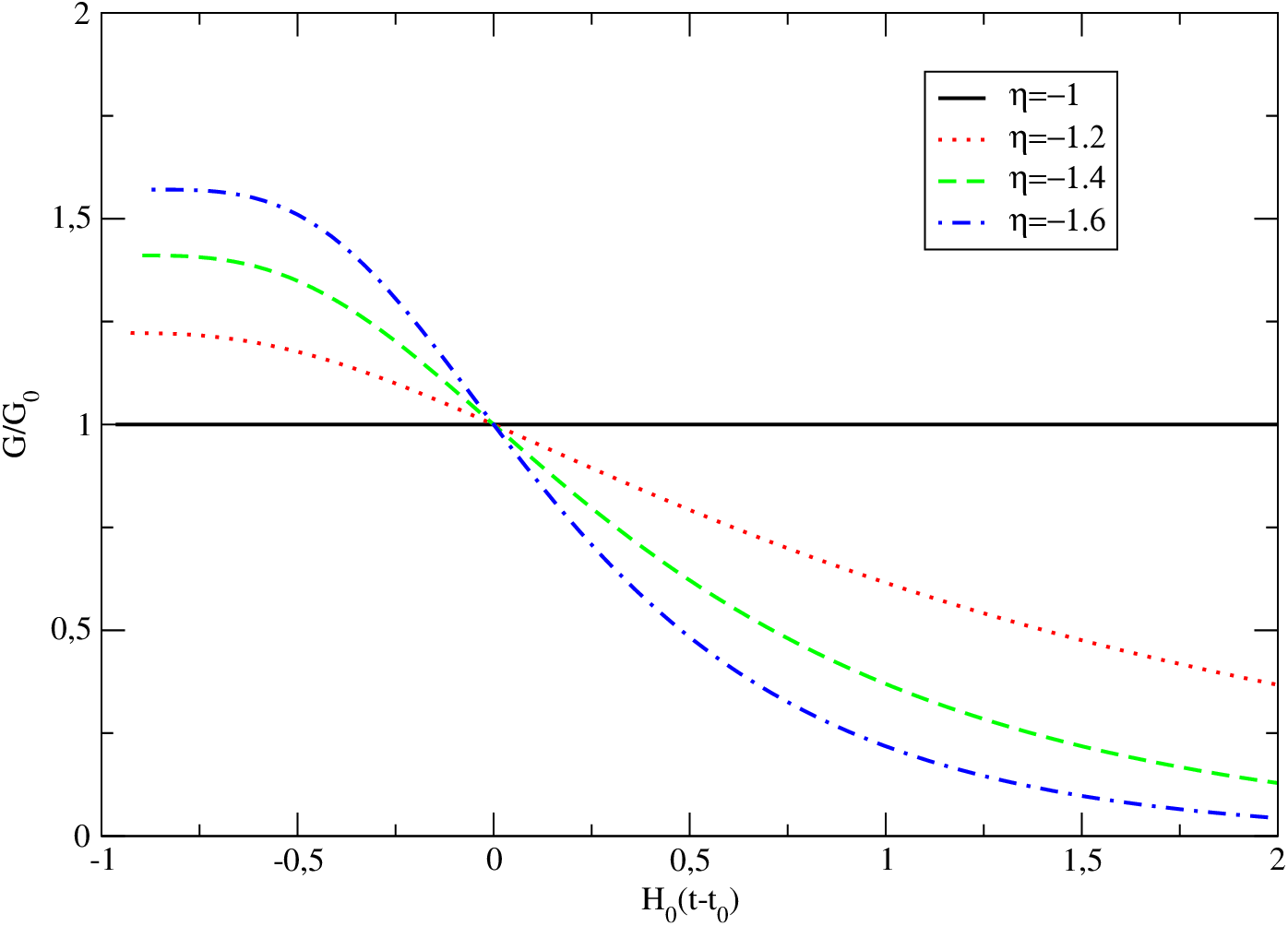}}}
\caption{\label{fig:G} The evolution of the gravitational constant $G$ is
given for $\Omega_{\Lambda}^{0} = 0.7$, $\gamma = 0$ and
four various values of the parameter $\eta$. In the very early
universe the change of $G$ is very slow, while in the distant future $G$ has a
tendency towards 0. For more negative values of the parameter $\eta$, 
the quantity $G$ changes faster.
}
\end{figure}

\begin{figure}
\centerline{\resizebox{1.0\textwidth}{!}{\includegraphics{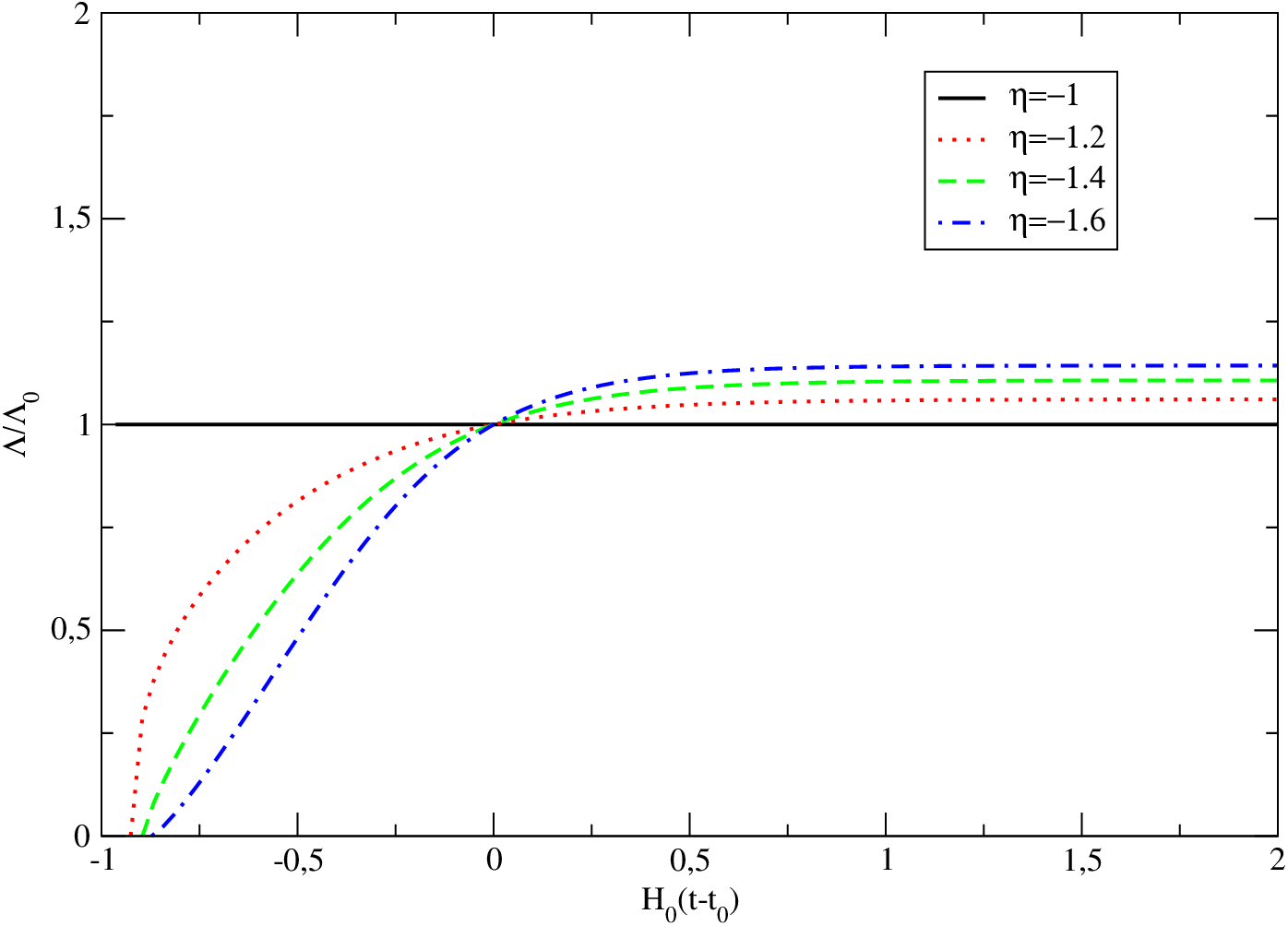}}}
\caption{\label{fig:Lambda} The change in time of the CC is depicted
 for $\Omega_{\Lambda}^{0} = 0.7$, $\gamma = 0$ and four different values of the 
 parameter $\eta$. In the distant future the quantity $\Lambda$ reaches a
 saturation value. The onset of saturation is sooner for the smaller values of
 the parameter $\eta$.
}
\end{figure}

\begin{figure}
\centerline{\resizebox{1.0\textwidth}{!}{\includegraphics{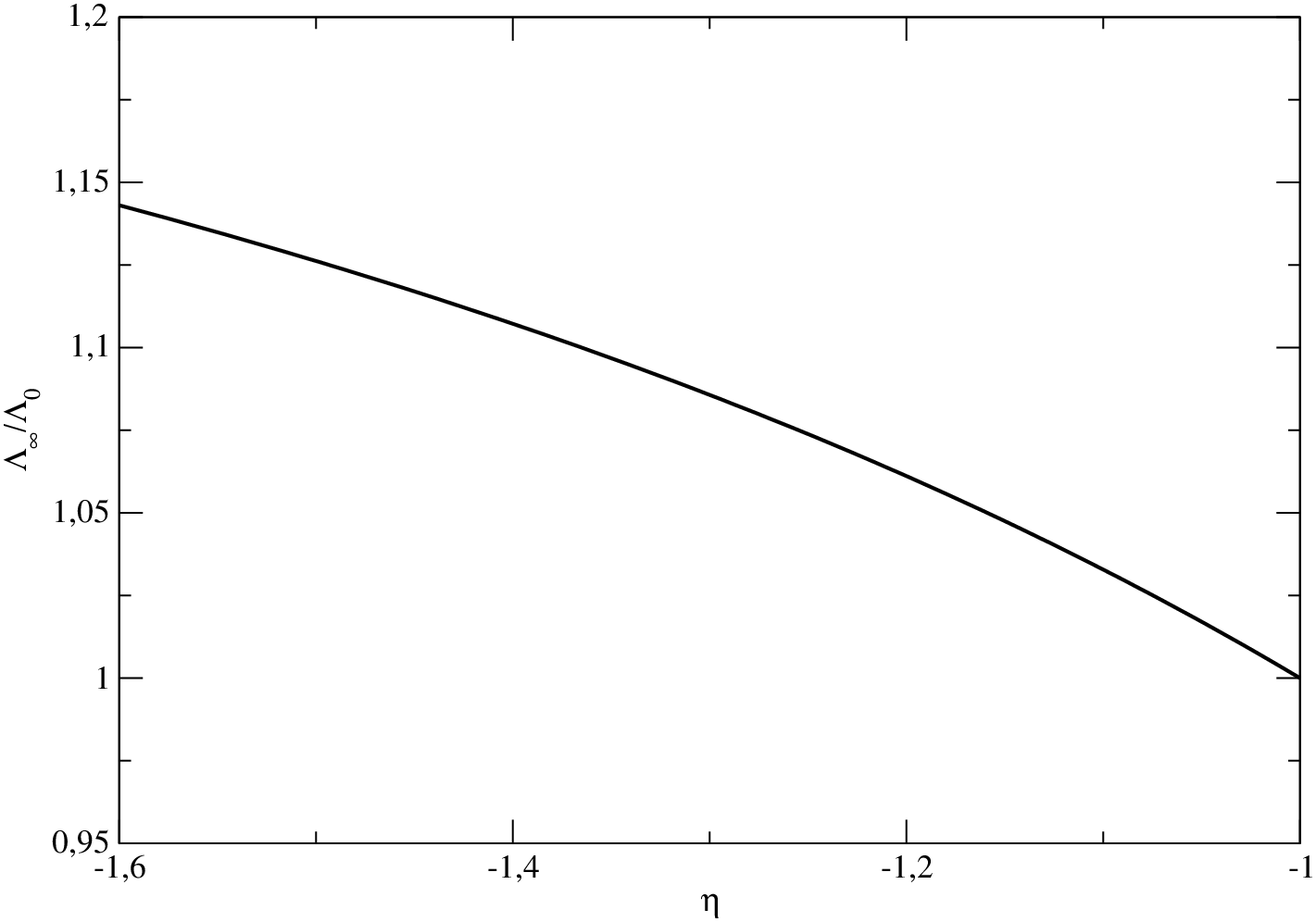}}}
\caption{\label{fig:w} The asymptotic value $\Lambda_{\infty}$ of the CC as a
function of the parameter $\eta$. Parameters used in the calculation are 
$\Omega_{\Lambda}^{0} = 0.7$, $\gamma = 0$. For a displayed range of values of the
parameter $\eta$, the asymptotic value $\Lambda_{\infty}$ is of the same order
of magnitude as the value of the CC today, $\Lambda_{0}$. }
\end{figure}

The evolution of the scale factor of the universe $a$ with time is shown in 
figure \ref{fig:a}. The dependences of $a$ on time for different values
of the parameter $\eta$ differ from each other the most in the distant 
past and the distant
future. Graphs of the evolution of the scale factor in the distant future 
also reveal the beginning of the exponential expansion, i.e. the onset of the de
Sitter regime. The time dependence of $G(t)$, depicted in figure
\ref{fig:G}, is more sensitive to the value of the parameter $\eta$.
Figure \ref{fig:G} displays some general features of the dynamics of 
$G(t)$. In the early universe (for small values of the scale factor) the change
of $G(t)$ is very slow. The rate of change is the greatest around the present
epoch, while at large times $G(t)$ tends towards 0.  
The pronounced dynamics of $G(t)$ for more negative values of the parameter
$\eta$ would probably be the best testing ground for the proposed class of
models, i.e. it would provide the most stringent constraint 
on the growth of the cosmological term
energy density with the scale factor, described by the parameter $\eta$. The
time evolution of $\Lambda$ is displayed in figure \ref{fig:Lambda}. In
the early universe the value of $\Lambda$ approaches $0$ for $\eta < -1$. 
For large times, the
function $\Lambda(t)$ tends to its asymptotic value $\Lambda_{\infty}$. For more
negative values of $\eta$, the asymptotic value $\Lambda_{\infty}$ increases and
it is approached slowlier.
The dependence of the asymptotic value $\Lambda_{\infty}$ on the parameter $\eta$ is
shown in figure \ref{fig:w}. For the chosen range of the parameter $\eta$,
approximately consistent with the constraint of reference \cite{odman}, the
quantity $\Lambda_{\infty}$ grows quite modestly, definitely remaining of the
same order of magnitude. 
 
The evolutions of $\Lambda(t)$ and $G(t)$ depicted in figures 
\ref{fig:G} and \ref{fig:Lambda}, respectively, show that the
universe ends up in a state with finite $\Lambda_{\infty}$ and vanishing
$G_{\infty}$. An important question for such a cosmology is the destiny of bound
systems. First, we focus on gravitationally bound systems. Let us consider a
spherically symmetric system with a mass $M$ at the centre, in the cosmology with
$\Lambda_{\infty} \approx const$ and $G_{\infty} \approx 0$. We can treat this
system as approximately static. The metrics of such a system is then given 
by \cite{ABS}
\begin{equation}
\label{eq:metrics}
d s^{2} = e^{\nu} d t^{2} - e^{\lambda} d r^{2} - r^{2}d \Omega^{2} \, ,
\end{equation} 
where
\begin{equation}
\label{eq:g00}
e^{\nu} = e^{-\lambda} = 1 - \frac{2 G_{\infty} M}{r} - \frac{1}{3} 
\Lambda_{\infty} r^{2} \, .
\end{equation}
In the Newtonian limit, the gravitational potential becomes
\begin{equation}
\label{eq:potential}
\phi(r) =  - \frac{G_{\infty} M}{r} - \frac{1}{6} \Lambda_{\infty} r^{2}\, .
\end{equation}  
As $G_{\infty}$ is vanishingly small and $\Lambda_{\infty}$ is finite, presently
gravitationally bound systems cannot remain bound. The unbounding happens at
some instance before a state with 
$\Lambda_{\infty} \approx const$ and $G_{\infty} 
\approx 0$ is achieved. Clusters of galaxies, galaxies, stars and stellar 
systems will fall apart. Planets will lose their atmospheres. The gravitational
interaction will become fully dominated by the asymptotic value of the
cosmological term.

Next, we turn to systems bound by forces other than gravitational, e.g. strong
or electromagnetic. In these systems, $G_{\infty}$ is clearly of no relevance. 
The dependence of $\Lambda_{\infty}$ on the parameter $\eta$ is shown in
figure \ref{fig:w}. For values of $\eta$ not much smaller than $-1$, $\Lambda_{\infty}$
remains of the same order of magnitude as $\Lambda_{0}$. As $\Lambda_{0}$
obviously has no unbounding effect on non-gravitationally bound systems, and
given the difference of many orders of magnitude 
between typical interaction scales
of gravitational and non-gravitational forces, neither $\Lambda_{\infty}$ will be
able to unbound non-gravitationally bound systems. The break-up of atoms and 
nucleons induced by the variable cosmological term will not happen. 

Finally, let us discuss constraints on the model parameters.
Cosmological observations impose rather strong constraints on the time 
dependence of the gravitational constant $G$. The observational bounds on the
variation of $G$ differ in their origin and refer to different epochs of the
universe expansion \cite{will, stairs}. The observations of spin-down rate of
pulsars yield constraints $|\dot{G}/G| \leq (2.2 - 5.5) \times 10^{-11} \rm
yr^{-1}$ for the pulsar PSR B0655+64 \cite{goldman}, and
$|\dot{G}/G| \leq (1.4 - 3.2) \times 10^{-11} \rm yr^{-1}$ for the pulsar
PSR J2019+2425 \cite{stairs,arzo}, where the range in constraints comes from
the uncertainties in the neutron star equation of state. 
The effects of the variation of $G$ on the orbital period of the 
neutron star-white dwarf binary system PSR B1855+09 give the constraint
$\dot{G}/G = (-1.3 \pm 2.7) \times 10^{-11} \rm yr^{-1}$ \cite{arzo,orbital}.
The consideration of effects of the variation of $G$ on the Chandrasekhar mass
in double-neutron-star binaries yields a constraint of
$\dot{G}/G = (-0.6 \pm 4.2) \times 10^{-12} \rm yr^{-1}$  \cite{binaries}.
The observations of SNIa give $\dot{G}/G \leq 10^{-11} \rm yr^{-1}$
at redshifts $z \sim 0.5$ \cite{snovae}, while the Big Bang Nucleosynthesis
constrains the gravitational constant during BBN to be $G_{BBN}/G_{0} = 1.01
\begin{array}{c} +0.20 \\ -0.16 \end{array}$ at the 68.3 \% confidence level  
\cite{copi}. One of the strongest constraints on the variation of $G$ comes from
the Lunar Laser Ranging (LLR) and amounts to 
$\dot{G}/G = (0.46 \pm 1.0) \times 10^{-12} \rm yr^{-1}$ \cite{LLR}. 
Figure \ref{fig:G} shows that constraints on $\dot{G}/G$ from the distant past
are more easily satisfied than those from the recent past since the variation of
$G$ is stronger in the recent than in the distant past. Namely, the rate of
change $\dot{G}$ grows with time in absolute value 
from the distant past to the present epoch,
while the value od $G$ decreases with time. For given $\eta$, of all past
epochs, the quantity $\dot{G}/G$ has the strongest variation for the present
epoch. On the other hand, the strongest
observational limit on $\dot{G}/G$ comes from LLR \cite{LLR} at the present
epoch (with $z=0$). 
Therefore, the comparison of theoretical expressions and observational
bounds at the present epoch 
can produce the most stringent constraints on the parameter $\eta$.
From equations (\ref{eq:Gevol2}) and (\ref{eq:varLambda}) the theoretical
expression for the rate $\dot{G}/G$ at $z=0$ becomes
\begin{equation}
\label{eq:limit}
\eta = -1 + \frac{1}{3 \Omega_{\Lambda}^{0} H_{0}} \left( \frac{\dot{G}}{G}
\right)_{0} \, .
\end{equation}
Using the value $H_{0}=72 \; km/s/Mpc$ for the Hubble parameter,
$\Omega_{\Lambda}^{0}=0.7$ and the
values for $\dot{G}/G$ from \cite{LLR}, we obtain the following constraints on
the parameter $\eta$: $\eta \ge -1.0035$ at the $1\sigma$ level, 
$\eta \ge -1.01$ at the $2\sigma$ level and 
$\eta \ge -1.016$ at the $3\sigma$ level. Other observational constraints 
yield weaker limits on $\eta$. The constraints from BBN \cite{copi}, e.g., using
(\ref{eq:Glaw}) yield $\eta \ge -1.1881$ at the 68.3 \% confidence level, which
is a much weaker limit. All the figures are plotted using a broader range of
parameter $\eta$ than the range constrained by LLR to better illustrate the form of
dependencies of functions $a$, $G$ and $\Lambda$ on time and $\eta$.
The obtained constraints
result in strong limits on the parameter $\eta$ which must be close to -1.
However, as long as the condition $\eta < -1$ is satisfied, no matter how close 
$\eta$ is to -1, the main conclusions of this paper remain unaltered. In another
words, even for very slow growth of $\rho_{\Lambda}$ (which satisfies all the
conditions on the variation of $G$) the partial rip scenario is valid, i.e., in
the distant future the gravitationally bound systems become unbound, while the
nongravitationally bound systems remain bound.

Considerations presented above show that a universe with the growing 
cosmological term and the
time-dependent $G$ has very interesting properties. The energy density of dark
energy component grows with the scale factor (like in phantom energy
models), but the parameter of the EOS remains -1 (unlike in phantom energy
models where the same parameter is less than -1). The universe ends up in a de
Sitter regime with constant $\Lambda_{\infty}$ and vanishing $G_{\infty}$. The
expansion continues infinitely, without any divergence of the 
scale factor in finite
time. However, despite the absence of the abrupt ending of the universe in a 
``Big Rip" event, accompanied by the unbounded growth of dark energy density,
 all bound systems do not remain bound. Gravitationally bound
systems become unbound in the distant future in an interplay of constant
repulsion of $\Lambda_{\infty}$ and vanishing attraction mediated by 
$G_{\infty}$. Systems bound by non-gravitational forces, however, do not share
the destiny of their gravitationally bound counterparts. Numerical calculations
show that, for a reasonable range of the parameter $\eta$, the value 
$\Lambda_{\infty}$ is not sufficiently large to disturb systems bound by, e.g.
strong or electromagnetic forces. It is important to emphasize once more
that all these findings are not dependent on the size of $\eta$, as long as it
remains less than -1. For the case of $\eta$ close to -1, all the effects
described above happen, only the variations of $G(t)$ and $\Lambda(t)$ are much
milder and the onset of the saturation of $\Lambda(t)$ in the future happens
much later.

Models with the growing cosmological term
 energy density and the time-de\-pen\-dent $G$ have
testable predictions about the past evolution of the universe. The confrontation
of these predictions with the host of cosmological observations available today
gives constraints on the parameters of the model, above all on the parameter
$\eta$. These constraints determine the destiny of the
universe and its structures and how far in the future the possible dramatic
unbounding effects lie in front of us.      

{\bf Acknowledgements.} The author would like to thank N. Bili\'{c}, B.Guberina 
and R. Horvat for useful comments on the manuscript. This work was supported by
the Ministry of Science and Technology of the Republic of Croatia under the 
contract No. 0098002.

\newpage

\newpage

\begin{thebibliography}{88}
\bibitem{Rev}  W.L. Freedman, M.S. Turner, Rev. Mod. Phys. 75 (2003) 1433;
 S.M. Carroll, astro-ph/0310342.
\bibitem{SN}  A.G. Riess et al., Astron. J. 116 (1998) 1009; S. Perlmutter et 
al., Astrophys. J. 517 (1999) 565.
\bibitem{CMBR} P. de Bernardis et al., Nature 404 (2000) 955; A.D. Miller et al.
Astrophys. J. Lett. 524 (1999) L1; S. Hanany et al., Astrophys. J. Lett. 545
(2000) L5; N.W. Halverson et al., Astrophys. J. 568 (2002) 38; B.S. Mason et
al., Astrophys. J. 591 (2003) 540; D.N. Spergel et al., Astrophys. J. Suppl. 
148 (2003) 175; L. Page et al. astro-ph/0302220.
\bibitem{wein} S. Weinberg, Rev. Mod. Phys. 61 (1989) 1.
\bibitem{peeb} P.J.E. Peebles, B. Ratra, Rev. Mod. Phys. 75 (2003) 559.
\bibitem{pad} T. Padmanabhan, Phys. Rept. 380 (2003) 235.
\bibitem{Q} B. Ratra, P.J.E. Peebles, Phys. Rev. D 37 (1988) 3406; 
P.J.E. Peebles, B. Ratra, Astrophys. J. 325 (1988) L17; C. Wetterich, Nucl.
Phys. B 302 (1988) 668; R.R. Caldwell, R. Dave, P.J. Steinhardt, Phys. Rev.
Lett. 80 (1998) 1582; I. Zlatev, L. Wang, P.J. Steinhardt, Phys Rev. Lett. 82
(1999) 896.
\bibitem{CG}  A. Yu. Kamenshchik, U. Moschella,  V. Pasquier, Phys. Lett. B511 
(2001) 265;  N. Bili\'{c}, G.B. Tupper, R.D. Viollier, Phys. Lett. B 535 (2002) 17;
 M.C. Bento, O. Bertolami, A.A. Sen, Phys. Rev. D 66 (2002) 043507. 
\bibitem{Freese}  K. Freese, M. Lewis, Phys. Lett. B 540 (2002) 1;  K. Freese,
Nucl. Phys. B (Proc. Suppl.) 124 (2003) 50.
\bibitem{Parker} L. Parker, A. Raval, Phys. Rev. D 60 (1999) 063512; 
Erratum-ibid. D67 (2003) 029901.
\bibitem{Ahmed} M. Ahmed, S. Dodelson, P.B. Greene, R. Sorkin, astro-ph/0209274.
\bibitem{mi}  A. Babi\'{c}, B. Guberina, R. Horvat, H. \v{S}tefan\v{c}i\'{c}, 
Phys. Rev. D 65 (2002) 085002; B. Guberina, R. Horvat, H. \v{S}tefan\v{c}i\'{c},
 Phys. Rev. D 67 (2003) 083001.
\bibitem{sola} I.L. Shapiro, J. Sola, Phys. Lett. B 475 (2000) 236; I.L. Shapiro, 
J. Sola, JHEP 0202 (2002) 006; I.L. Shapiro, J. Sola, C. Espana-Bonet, 
P. Ruiz-Lapuente, Phys. Lett. B 574 (2003) 149.
\bibitem{bonanno} A. Bonanno, M. Reuter, Phys. Lett. B 527 (2002) 9; 
E. Bentivegna, A. Bonanno, M. Reuter, JCAP 0401 (2004) 001.
\bibitem{odman} A. Melchiorri, L. Mersini, C.J. \"{O}dman, M. Trodden, Phys.
Rev. D 68 (2003) 043509.
\bibitem{cald} R.R. Caldwell, Phys. Lett. B 545 (2002) 23.
\bibitem{phantom} C. Armend\'{a}riz-Pic\'{o}n, T. Damour, V. Mukhanov, Phys. Lett.
 B 458 (1999) 209; T. Chiba, T. Okabe, M. Yamaguchi, Phys. Rev. D 62 (2000) 
 023511; V. Faraoni, Int. J. Mod. Phys. D 11 (2002) 471; S.M. Carroll, 
 M. Hoffman, M. Trodden, Phys. Rev. D 68 (2003) 023509; S. Nojiri, S.D.
 Odintsov, Phys. Lett. B 562 (2003) 147; S. Nojiri, S.D. Odintsov, Phys. Lett 
 B 565 (2003) 1; P. Singh, M. Sami, N. Dadhich, Phys. Rev. D 68 (2003) 023522;
 G.W. Gibbons, hep-th/0302199; L.P.Chimento, R. Lazkoz, Phys. Rev. Lett. 91 
 (2003) 211301; J.G. Hao,
 X.Z. Li, Phys. Rev. D 68 (2003) 083514; M.P.Dabrowski, T. Stachowiak, M.
 Szydlowski, Phys.Rev. D68 (2003) 103519; 
 J.G. Hao, X.Z. Li Phys. Rev. D 67 (2003) 107303;
 J.G. Hao, X.Z. Li, astro-ph/0309746.  
\bibitem{cald2}  R.R. Caldwell, M. Kamionkowski, N.N. Weinberg, Phys. Rev. Lett. 
91 (2003) 071301.
\bibitem{Gvar} A. Beesham, Nuovo Cimento 96 B (1986) 17; A. Beesham, Int. J.
Theor. Phys. 25 (1986) 1295; A-M.M. Abdel-Rahman, Gen. Rel. Grav. 22 (1990) 655;  
M.S. Berman, Phys. Rev. D 43 (1991) 1075; M.S.
Berman, Gen. Rel. Grav. 23 (1991) 465; R.F. Sistero, Gen. Rel. Grav. 32 (1991)
1265; D. Kalligas, P. Wesson, C.W.F. Everitt, Gen. Rel. Grav. 24 (1992) 351; 
T. Singh, A. Beesham, Gen. Rel. Grav. 32 (2000) 607; A.I. Arbab, 
A. Beesham, Gen. Rel. Grav. 32 (2000) 615; A.I. Arbab, Spacetime and Substance 
1(6) (2001) 39; A.I. Arbab, astro-ph/0308068; 
J. Ponce de Leon, Class. Quant. Grav. 20 (2003) 5321.
\bibitem{stef} H. \v{S}tefan\v{c}i\'{c}, astro-ph/0310904.
\bibitem{Odin} E. Elizalde, S.D. Odintsov, Phys. Lett. B 303 (1993) 240; E.
Elizalde, S.D. Odintsov, Phys. Lett. B 321 (1994) 199; E. Elizalde, S.D.
Odintsov, I.L. Shapiro, Class. Quant. Grav. 11 (1994) 1607; E. Elizalde, C.
Lousto, S.D. Odintsov, A. Romeo, Phys. Rev. D 52 (1995) 2202.
\bibitem{FAFM} K. Freese, F.C. Adams, J.A. Frieman, E. Mottola, Nucl. Phys. B
287 (1987) 797.
\bibitem{ABS} see e.g. R. Adler, M. Bazin, M. Schiffer, Introduction to general
relativity, McGraw-Hill, 1975.
\bibitem{will} C.M. Will, Living Rev. Rel. 4 (2001) 4. 
\bibitem{stairs} I.H. Stairs, Living Rev. Rel. 6 (2003) 5.
\bibitem{goldman} I. Goldman, Mon. Not. Roy. Astr. Soc. 244 (1990) 184.
\bibitem{arzo} Z. Arzoumanian, PhD thesis, Princeton University, Princeton,
N.J., U.S.A., 1995.
\bibitem{orbital} V.M. Kaspi, J.H. Taylor, M. Ryba, Astrophys. J. 428 (1994)
713.
\bibitem{binaries} S.E. Thorsett, Phys. Rev. Lett. 77 (1996) 1432.
\bibitem{snovae} E. Gaztanaga, E. Garcia-Berro, J.Isern, E. Bravo, 
I. Dominguez, Phys. Rev. D 65 (2002) 023506.
\bibitem{copi}  C.J. Copi, A.N. Davis, L.M. Krauss, astro-ph/0311334.
\bibitem{LLR}  S.G. Turyshev, J.G. Williams, K. Nordtvedt Jr., M. Shao, T.W.
Murphy Jr., gr-qc/0311039. 
\end{thebibliography}
\end{document}